\documentclass[conference]{IEEEtran}
\usepackage{url}
\usepackage{cite}
\usepackage{amsmath,amssymb,amsfonts}
\usepackage{algorithm}
\usepackage{algorithmic}
\usepackage{graphicx}
\usepackage{textcomp}
\usepackage{xcolor}
\usepackage{comment}

\begin{document}
\title{Efficient executions of Pipelined Conjugate Gradient Method on Heterogeneous Architectures}

\author{\IEEEauthorblockN{Manasi Tiwari}
\IEEEauthorblockA{Department of Computational and Data Sciences\\
Indian Institute of Science, Bangalore\\
Email: manasitiwari@iisc.ac.in}
\and
\IEEEauthorblockN{Sathish Vadhiyar}
\IEEEauthorblockA{Department of Computational and Data Sciences\\ Indian Institute of Science, Bangalore\\
Email: vss@iisc.ac.in}}

\maketitle

\begin{abstract}
The Preconditioned Conjugate Gradient (PCG) method is widely used for solving linear systems of equations with sparse matrices. A recent version of PCG, Pipelined PCG, eliminates the dependencies in the computations of the PCG algorithm so that the non-dependent computations can be overlapped with communication.
In this paper, we propose three methods for efficient execution of the Pipelined PCG algorithm on GPU accelerated heterogeneous architectures. 

The first two methods achieve task-parallelism using asynchronous executions of different tasks on CPU cores and GPU. The third method achieves data parallelism by decomposing the workload between CPU and GPU based on a performance model.
The performance model takes into account the relative performance of CPU cores and GPU using some initial executions and performs 2D data decomposition. 
We also implement optimization strategies like kernel fusion for GPU and merging vector operations for CPU. Our methods give up to 8x speedup and on average 3x speedup over PCG CPU implementation of Paralution and PETSc libraries.  They also give up to 5x speedup and on average 1.45x speedup over PCG GPU implementation of Paralution and PETSc libraries. The third method also provides an efficient solution for solving problems that cannot be fit into the GPU memory and gives up to 2.5x speedup for such problems. 
\end{abstract}


\section{Introduction}
Many High Performance Computing applications in Computational Fluid Dynamics, Electromagnetics, Finance etc. need to solve Partial Differential Equations over space and time. These partial differential equations are discretized using finite volume, finite element or finite difference methods and they result in a linear system of equations $Ax = b$. Generally, the $A$ matrix obtained by using these discretization schemes is large and sparse. Iterative methods are used to solve these linear systems of equations with large sparse matrices. In iterative methods, we begin with an initial guess $x_0$ and iterate till we get a correct solution subject to an error tolerance defined by us. The most widely used iterative methods for solving these sparse systems are Krylov Subspace methods. 
The basic idea behind Krylov methods when solving a linear system $Ax = b$ is to build a solution within the Krylov subspace composed of several powers of matrix A multiplied by vector b, that is, $\{b,Ab,A^2b, ...,A^mb\}$.

Conjugate Gradient (CG) \cite{Hestenes&Stiefel:1952} \cite{10.5555/829576} method is one of the most widely used Krylov Subspace methods. It is used to find the solution of linear systems with symmetric sparse positive definite matrices.
In exact arithmetic, it gives the solution of a system of size N in N steps.
A preconditioner can be applied to the system to condition the input system and to improve convergence.

Today's HPC systems are heterogeneous. They have different processing entities along with traditional processors. Accelerators like Intel Xeon Phi, FPGAs, and Nvidia's General Purpose Graphics Processing Units (GP-GPUs) are examples of such entities. The programming models for these different accelerators are different from that of the traditional processors as well.
In order to use all the resources available within a node and across multiple nodes efficiently, we must interleave the features in the programming models for these different processing entities in such a way that we achieve the best possible performance from the platform. 

The main computational kernels in Preconditioned Conjugate Gradient (PCG) method are Sparse Matrix Vector Product (SPMV), Preconditioner Application (PC), Vector-Multiply-Adds (VMAs) and Dot Products.
In order to improve the performance of the PCG solver on GPU, the existing  works have concentrated on improving performance of the most time consuming kernels, i.e. PC and SPMV kernels\cite{10.1007/978-3-319-17248-4_4}\cite{Naumov11incomplete-luand}\cite{7013063}\cite{article}\cite{5481803} but none of these try to improve the performance of the PCG solver as a whole using CPU cores and GPU.

For distributed memory systems, the bottleneck in PCG is the synchronization that happens across all cores due to the allreduce operations from the dot products in the algorithm.
Hence, existing research has worked on reducing the number of allreduces to one per iteration as opposed to the three that exists in the na\"ive algorithm.
CG methods with one allreduce per iteration have been presented in the earlier works \cite{10.5555/829576} \cite {10.5555/898717} \cite{10.1016/0377-0427(89)90045-9}.

With the advent of non-blocking collectives like MPI\_IAllreduce in the MPI-3 standard \cite{mpich}, the overlapping of allreduce with useful work has been made possible.
Pipelined PCG (PIPECG) proposed by Ghysels et al\cite{10.1016/j.parco.2013.06.001}, on which this work is based, uses Chronopoulos-Gear PCG \cite{10.1016/0377-0427(89)90045-9} which has one allreduce per iteration. By introducing extra VMAs, they remove the dependencies between the dot products and PC+SPMV and then they overlap the allreduce with the PC and SPMV. The resulting algorithm offers another advantage which makes it a perfect candidate for our hybrid implementations. As the PC and SPMV in PIPECG don't depend on results of the previous dot products, we can execute them simultaneously on multi-core CPU and GPU in heterogeneous architecture. This would require communicating data between CPU and GPU, thus introducing additional costs. We show that by using asynchronous streams efficiently, we can hide the complete time for data movement between CPU and GPU.

We propose three methods for efficient execution of PIPECG on GPU accelerated architecture.
The first two methods, \textbf{Hybrid-PIPECG-1} and \textbf{Hybrid-PIPECG-2}, achieve task-parallelism using asynchronous executions of different tasks on multi-core CPU and GPU. The third method, \textbf{Hybrid-PIPECG-3}, achieves data parallelism by decomposing the workload between multi-core CPU and GPU based on a performance model. The model takes into account the relative performance of each processing entity in heterogeneous architecture using some initial executions and decomposes the data according to these performances. We use CUDA streams for asynchronous data movements between CPU and GPU. We implement optimization strategies like kernel fusion for GPU and merged vector operations for CPU cores. Our methods give up to 8x speedup and on average 3x speedup over PCG CPU implementation of Paralution and PETSc libraries.  Our methods give up to 5x speedup and on average 1.45x speedup over PCG GPU implementation of Paralution and PETSc libraries. Hybrid-PIPECG-3 method also provides an efficient solution for solving problems that cannot be fit into the GPU memory and gives up to 2.5x speedup for such problems. 

The rest of the paper is organized as follows: Section \ref{relatedwork} gives the related work, Section \ref{back} gives background related to PCG and PIPECG, Section \ref{method} describes our three proposed methods.
Section \ref{optimization} presents the implementation details and optimizations we have used in our methods. Section \ref{experiment} presents experiments, results and discussions for our proposed methods and Section \ref{conclusion} gives the conclusions and future work.

\section{Related Work}
\label{relatedwork}

To achieve optimum performance of the PCG method on GPU accelerated heterogeneous architectures, many works have concentrated on efficient GPU implementation of the PC kernel. Algebraic Multigrid GPU implementations are presented in \cite{doi:10.1137/110838844}\cite{10.1007/978-3-319-17248-4_4}. Incomplete LU and cholesky factorizations on GPUs are presented in \cite{Naumov11incomplete-luand} \cite{10.1007/s11227-012-0825-3}.
Research works also concentrate on optimizing the most time consuming kernel in PCG, the SPMV kernel \cite{10.1145/1654059.1654078}. Different sparse matrix formats have been proposed in \cite{article} to improve SPMV performance on GPUs. Optimization techniques like kernel fusion and using texture memory for CG solvers on GPUs were presented in \cite{5481803}\cite{10.1145/2907944}.
All the works mentioned above concentrate on kernel executions and optimizations only on the GPUs. They do not utilize the multi-core CPU also present in the system.
Our work is different from all the works described above since our work aims to utilize all the available resources of the system and accelerate the performance of the PCG method as a whole.

The authors in \cite{https://doi.org/10.1002/fld.2462} present a technique to use the multi-core CPUs to perform SPMV computations of the off-diagonal elements when the CG method is implemented on multiple GPUs. They execute only SPMV computations on the CPU. This is beneficial only for a small range of structured matrices. Our work is different from this work as we use the multi-core CPU for every operation rather than just SPMV. This gives benefits for all kinds of matrices. To the best of our knowledge, ours is the first work to recognize the advantage of independent computations in PIPECG method for heterogeneous architectures and present its hybrid implementations.

\section{Background}
\label{back}

\subsection{PCG:}The Preconditioned Conjugate Gradient Method (PCG) introduced by Hestenes and Stiefel\cite{Hestenes&Stiefel:1952} is given in Algorithm \ref{pcg_algo}.
\begin{algorithm}
\caption{Preconditioned Conjugate Gradient (PCG)}
\begin{algorithmic}[1]
      \STATE $r_0 = b-Ax_0;$ $u_0 = M^{-1}r_0;$ 
      \STATE $\gamma_0 = (u_0, r_0);$ $norm_0 = \sqrt{(u_0,u_0)}$
      \FOR{i=0,1...}
      \IF{$i>0$} 
      \STATE $\beta_i=\gamma_i/\gamma_{i-1}$
      \ELSE
      \STATE $\beta_i=0$
      \ENDIF
      \STATE $p_{i} = u_i + \beta_i p_{i-1}$
      \STATE $s = Ap_i$
      \STATE $\delta = (s,p_i)$
      \STATE $\alpha = \gamma_i/\delta$
      \STATE $x_{i+1} = x_i + \alpha p_i$
      \STATE $r_{i+1} = r_i - \alpha s$
      \STATE $u_{i+1} = M^{-1}r_{i+1}$
      \STATE $\gamma_{i+1} = (u_{i+1}, r_{i+1});$ 
      \STATE$norm_{i+1} = \sqrt{(u_{i+1},u_{i+1})}$
      \ENDFOR
\end{algorithmic}
\label{pcg_algo}
\end{algorithm}
As shown in Algorithm \ref{pcg_algo}, the computational kernels in PCG are Sparse Matrix Vector Product (SPMV) in line 10, Preconditioner Application (PC) in line 15, Vector-Multiply-Adds (VMAs) in lines 13 and 14 and dot products in lines 11, 16 and 17. Note that the third dot product (line 17) is for calculating the residual norm which is used to check for convergence.

In the na\"ive PCG Algorithm \ref{pcg_algo}, we can see that the operation in every line depends on the operation in the previous line. There are no independent computations in each iteration of PCG which can be executed simultaneously.  

\subsection{PIPECG:}
The Pipelined Preconditioned Conjugate Gradient Method (PIPECG) was proposed by Ghysels and Vanroose\cite{10.1016/j.parco.2013.06.001} for obtaining performance improvements on distributed memory architectures.
\begin{algorithm}
\caption{Pipelined Preconditioned Conjugate Gradient (PIPECG)}
\begin{algorithmic}[1]
      \STATE $r_0 = b-Ax_0;$ $u_0 = M^{-1}r_0;$ $w_0 = Au_0;$
      \STATE $\gamma_0 = (r_0, u_0);$ $\delta = (w_0, u_0);$ $norm_0 = \sqrt{(u_0, u_0)}$
      \STATE $m_0 = M^{-1}w_0;$ $n_0 = Am_0$
      \FOR{i=0,1...}
      \IF{$i>0$} 
      \STATE $\beta_i=\gamma_i/\gamma_{i-1};$ $\alpha_i=\gamma_i/(\delta - \beta_i\gamma_i/alpha_{i-1});$
      \ELSE
      \STATE $\beta_i=0; \alpha_i = \gamma_i/\delta$
      \ENDIF
      \STATE $z_{i} = n_i + \beta_i z_{i-1} $ 
      \STATE$q_{i} = m_i + \beta_i q_{i-1}$ 
      \STATE $s_{i} = w_i + \beta_i s_{i-1}$
      \STATE $p_{i} = u_i + \beta_i p_{i-1}$ 
      \STATE $x_{i+1} = x_i + \alpha_i p_{i}$ 
      \STATE $r_{i+1} = r_i - \alpha_i s_{i}$
      \STATE $u_{i+1} = u_i - \alpha_i q_{i}$ 
      \STATE$w_{i+1} = w_i - \alpha_i z_{i}$
      \STATE $\gamma_{i+1} = (r_{i+1}, u_{i+1})$ 
      \STATE $\delta = (w_{i+1}, u_{i+1})$ 
      \STATE $norm_{i+1} = \sqrt{(u_{i+1}, u_{i+1})}$
      \STATE $m_{i+1} = M^{-1}w_{i+1}$
      \STATE $n_{i+1} = Am_{i+1}$ 
      \ENDFOR
\end{algorithmic}
\label{pipelined_pcg_algo}
\end{algorithm}
As shown in Algorithm \ref{pipelined_pcg_algo}, PIPECG introduces extra VMA operations (on lines 10,11,12,16,17) to remove the dependencies between the dot products (lines 17,18,19) and PC (line 21) and SPMV (line 22) so that PC and SPMV can be computed while dot products are being computed.

Even though PIPECG method was proposed for distributed memory architecture, we use it in our work because of the independent computations it provides.
The dot products can be executed on the CPU while PC+SPMV can be executed simultaneously on GPU as they are not dependent on each other. This strategy helps us to utilize all the resources in the GPU accelerated node and achieve optimum performance.

\subsection{CUDA Streams} Programs to be run on Nvidia's GPUs are written using Compute Unified Device Architecture (CUDA) programming model. All the CUDA kernels and functions are launched on the default stream. Nvidia introduced user defined CUDA streams from the Kepler GPU architecture series. With user defined streams, we can launch these kernels in different streams to achieve better concurrency. The streams allow overlap of CPU-GPU data movements with the GPU kernel executions. We use CUDA streams for implementing PIPECG algorithm on GPU accelerated architectures.

\section{Methodology}
\label{method}

\subsection{Hybrid-PIPECG-1 Method}
\label{hybrid1method}
In the GPU implementation of the na\"ive PCG, the CPU launches CUDA kernels for VMAs, dot products, PC and SPMV on the GPU and then remains idle. Furthermore, on the GPU, these kernels are executed one after the other since they are dependent on the previous kernel for result. 

With the PIPECG method, we have independent kernels and thus we can make use of the idle CPU. We can achieve task parallelism by asynchronous executions of independent tasks on CPU and GPU.

\begin{figure}[htbp]
\includegraphics[width=\columnwidth]{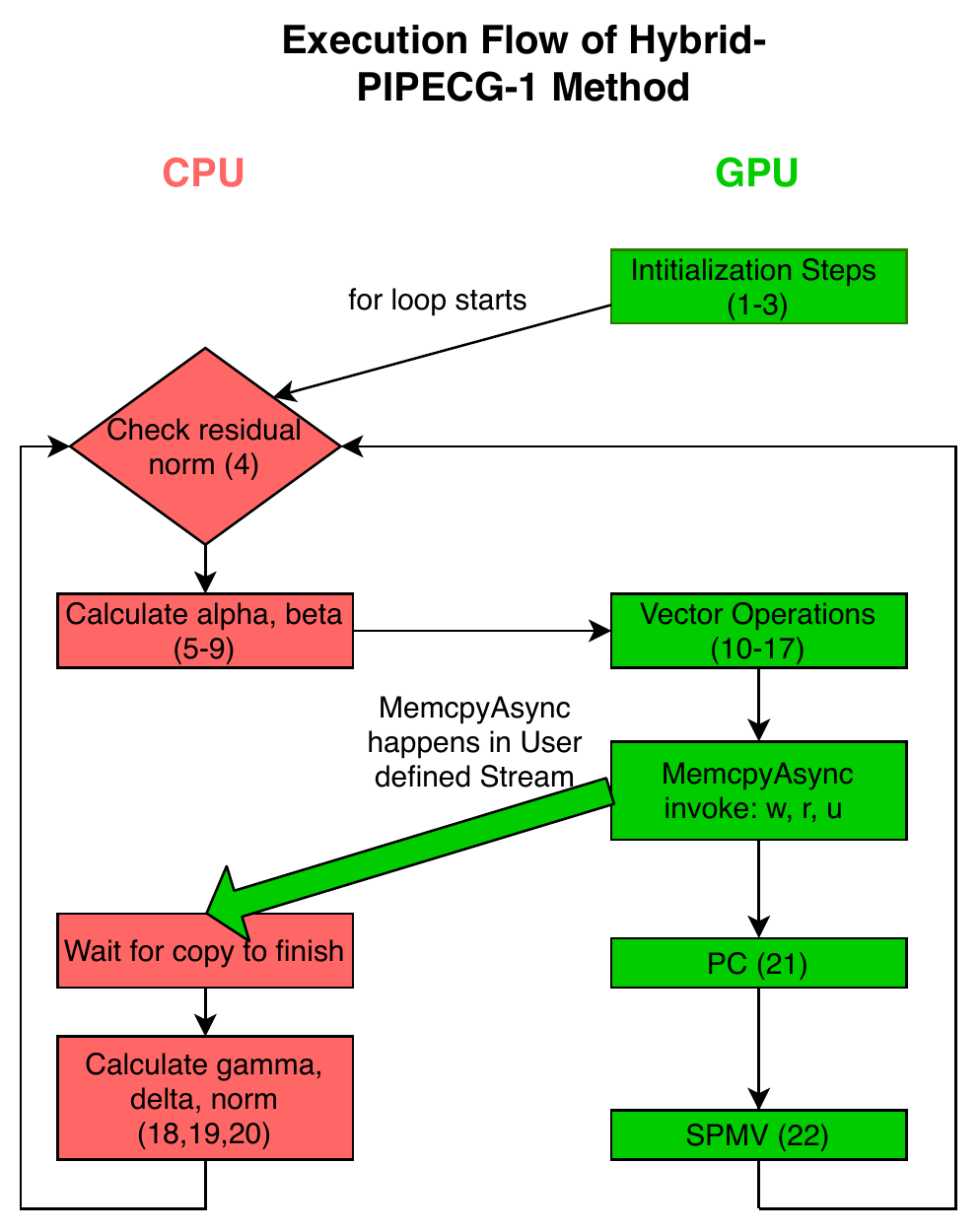}
\caption{Execution flow of Hybrid-PIPECG-1 method}
\label{hybrid1fig}
\end{figure}

We show the execution flow of the proposed Hybrid-PIPECG-1 method in Figure \ref{hybrid1fig}. Note that the matrix $A$, the vectors $b$ and $x$ have been moved to the GPU prior to this execution flow. The number of rows in $A$ is N and the number of non-zeroes in $A$ are nnz. 

The execution flow has two columns. They show the operations performed on the CPU and those performed on the GPU. The rectangular boxes show the operation performed and the number within the bracket is the line number of Algorithm \ref{pipelined_pcg_algo} that the box executes. The condition box checks the residual norm against the specified tolerance. The solid thick arrow represents data movement and its direction shows the source and destination of the data movement.

The implementation starts with executing the initialization steps (lines 1-3 of Algorithm \ref{pipelined_pcg_algo}) on the GPU. After this, the for loop starts which iterates until the norm of the preconditioned residual becomes smaller than the user defined tolerance. In each iteration, first $\alpha$ and $\beta$ are calculated on the CPU. Then the Vector Operations (lines 10-17) are executed on the GPU which update the vectors $w$, $r$ and $u$ among others. 

We know that dot products $\gamma$, $\delta$ and residual $norm$ on lines 18,19,20 can be executed simultaneously with PC and SPMV on lines 21 and 22.

For executing dot products $\gamma$, $\delta$ and residual $norm$ on the CPU, the CPU needs to have the vectors $w$, $u$ and $r$ but as the updated vectors are on the GPU, we have to copy them to the CPU at every iteration. So here, we define a stream which asynchronously copies $w$, $r$ and $u$ while GPU carries on with its kernel executions. The CPU waits on this stream till the copy is finished and then proceeds to calculate $\gamma$, $\delta$ and $norm$. 

Thus, in Hybrid-PIPECG-1, PC and SPMV computations on the GPU are overlapped with the data movement from GPU to CPU and the dot product calculation on the CPU.

\subsection{Hybrid-PIPECG-2 Method}
\label{hybrid1method}

The Hybrid-PIPECG-1 method described in the previous subsection requires to copy three vectors from GPU to CPU in every iteration. Since 3N elements are copied from GPU to CPU, it can become costly for linear systems of equations with vectors of large size N, as the time for copying will exceed the time for GPU kernel executions thus degrading the overall performance. We develop Hybrid-PIPECG-2 method to reduce the number of vectors to be copied from GPU to CPU in every iteration. This is achieved by performing some redundant computations on CPU and GPU. 

If we want to compute the dot products $\gamma$, $\delta$ and $norm$ on the CPU, we need to have $w$, $u$ and $r$ vectors on the CPU. Instead of copying the updated vectors from the GPU at every iteration, we can update them on the CPU itself. In the PIPECG method of Algorithm \ref{pipelined_pcg_algo}, we see that we can update $w$, $u$ and $r$ on the CPU using the vectors $z$, $q$ and $s$. In turn, we would need $n$ and $m$ for updating $z$ and $q$. This means the CPU should have a copy of $z$, $q$, $s$, $n$, $m$ in addition to $w$, $u$ and $r$. For updating these vectors, we would need to copy only $n$ vector from the GPU to CPU. All the other vectors can be updated on the CPU itself. Note that all these updates happen on the full vectors of size N. 

\begin{figure}[htbp]
\includegraphics[width=\columnwidth]{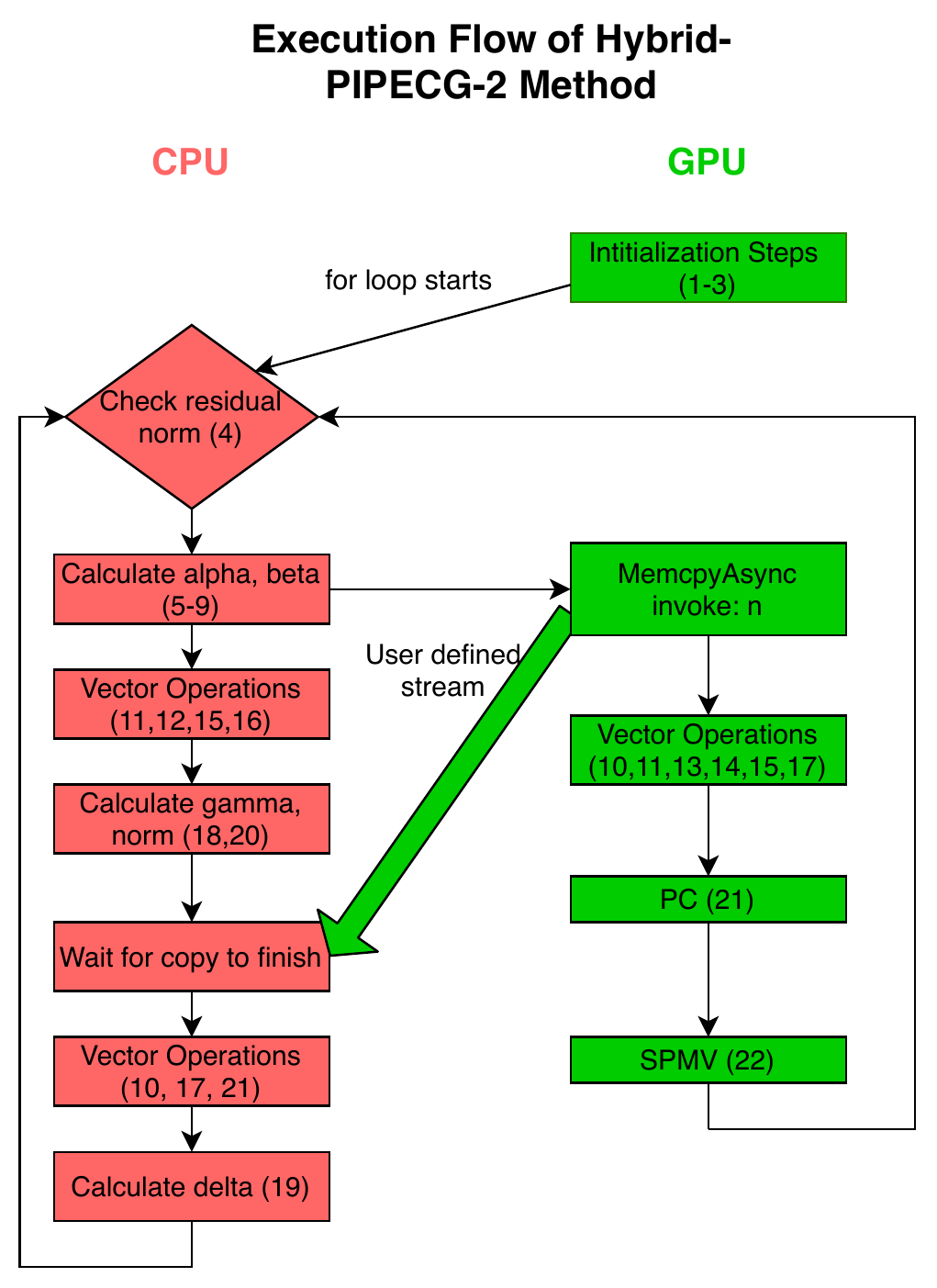}
\caption{Execution flow of Hybrid-PIPECG-2 method}
\label{Hybrid2fig}
\end{figure}

Figure \ref{Hybrid2fig} shows the execution flow of the Hybrid-PIPECG-2 method. In the for loop, after calculating $\alpha$ and $\beta$, the vector $n$ is copied from the GPU to the CPU on the user defined stream. While the copy is progressing, both CPU and GPU perform their operations. GPU proceeds with its Vector Operations, PC and SPMV kernels. On the CPU, we observe that for updating the vectors $z$, $w$ and $m$, CPU needs the vector $n$. While $n$ is being copied, CPU can proceed with the update of vectors $q$, $s$, $r$ and $u$ as they don't need $n$. After vector updates, $\gamma$ and $norm$ can be calculated. Then the CPU waits on the user defined stream until the copy is finished. After $n$ is successfully received, CPU can proceed to update $z$, $w$ and $m$ vectors. 
After the vector updates, $\delta$ can be computed.

Thus, with Hybrid-PIPECG-2 method, we are able to reduce the number of vector copies to one per iteration. We copy N elements from GPU to CPU every iteration. This is done at the cost of redundant computations because we need to update vectors $z$, $q$, $u$, $w$ and $m$ on both CPU and GPU.
Moreover, the data movement is hidden by computations on the CPU so the CPU doesn't have to be idle while the copy is happening.

\subsection{Hybrid-PIPECG-3 Method}
\label{hybrid3method}
The Hybrid-PIPECG-1 and Hybrid-PIPECG-2 methods achieve task parallelism by executing independent kernels on CPU and GPU simultaneously and they achieve asynchronous copy through streams to hide the data movement. But for linear systems with large N, the extra VMAs introduced by PIPECG algorithm in order to eliminate the dependencies between computations contribute to a great overhead. They increase the cost of each iteration of PIPECG and the benefit of simultaneous computations on CPU and GPU is overshadowed by this overhead.

Additionally for large N, executing redundant computations for full vectors of length N on both CPU and GPU proves to be counter-productive. This motivates us to develop a data parallel method, where we divide the matrix and vectors between the CPU and GPU, so that they perform operations on their own parts and communicate data when needed.

This motivates us to develop a data parallel PIPECG method, where we divide the matrix and vectors between the CPU and GPU, so that they perform operations on their own parts and communicate data when needed. We develop Hybrid-PIPECG-2 where we calculate the relative performance of CPU cores and GPU and decompose the data between the CPU and GPU using these performances. We also implement 2D decomposition of the data for better overlap of computations with communication. The method consists of 3 parts: Performance Modelling, Data Decomposition, and the actual PIPECG iterations.

\subsubsection{Performance Modelling}
\label{performance_modelling}
The times taken by the CPU and GPU to perform the same operation on the same amount of data can be different. Hence, when we want to decompose data between the CPU and GPU, we should make sure that the amount of data given to each one of them is according to their processing power. This will ensure that a situation where one of them is idle and the other is working never arises.
In order to achieve this decomposition, first we need to have a measure of the relative performance of the CPU and GPU. 

For this, we execute the SPMV kernel for the full matrix $A$ (nnz elements) on CPU and GPU separately. We choose the SPMV kernel because that is the most time dominating kernel in the PIPECG iteration. If we decompose the data in a way such that the time taken by CPU for SPMV kernel on its data is equal to the time taken by GPU for the SPMV kernel on its data, then complete overlap of the most time consuming kernel is achieved. Hence, we perform five executions of SPMV on both CPU and the GPU for nnz elements.  We perform five executions so that effects of cache locality that become prevalent in the later iterations are also be taken into consideration.

Once we have the time taken by CPU cores, $t_{cpu}$ and the time taken by GPU, $t_{gpu}$, we calculate the performance of cores CPU, $s_{cpu}$ and the performance of GPU, $s_{gpu}$ as follows-\\
$s_{cpu}=nnz/t_{cpu}$\\
$s_{gpu}=nnz/t_{gpu}$\\
Then, we calculate the relative performance $r_{cpu}$ and $r_{gpu}$ as follows-\\
$r_{cpu}=s_{cpu}/(s_{cpu}+s_{gpu})$\\
$r_{gpu}=s_{gpu}/(s_{cpu}+s_{gpu})$\\
After we obtain $r_{cpu}$ and $r_{gpu}$, we now divide the $nnz$ into two parts, $nnz_{cpu}$ and $nnz_{gpu}$ as follows-\\
$nnz_{cpu}=nnz*r_{cpu}$\\
$nnz_{gpu}=nnz - nnz_{cpu}$

For ease of implementation, we do not assign exact $nnz_{cpu}$ elements to CPU and $nnz_{gpu}$ elements to GPU. Instead, we find out the number of rows to be assigned to the CPU, $N_{cpu}$, which would contain either equal to or slightly less number of non-zeroes than $nnz_{cpu}$. This gives a 1-D decomposition of the $A$ matrix. $N_{gpu}$ is then obtained by $N - N_{cpu}$.

\subsubsection{Data Decomposition}
\label{data_decomposition}
Now that we have $N_{cpu}$ and $N_{gpu}$, we easily assign $N_{cpu}$ number of rows to the CPU and $N_{gpu}$ number of rows to the GPU. We also divide the vectors between the CPU and GPU using same parameters. We give $N_{cpu}$ elements of each vector to the CPU and the other $N_{gpu}$ elements to the GPU. 

The division of vectors ensures that there are no redundant computations as both CPU and GPU will be acting on just their local elements. But in every iteration, the SPMV kernels of both CPU and GPU will require the full $m$ vector. After 1-D decomposition, the CPU has $N_{cpu}$ elements of the $m$ vector and the GPU has the other $N_{gpu}$ elements. It is clear that we need to copy these partial vectors from their home device to the other device. 

In order to hide the time taken for this copy, we perform a further decomposition of the $nnz_{cpu}$ into $nnz1_{cpu}$ and $nnz2_{cpu}$ in such a way that all the nnz's in $nnz1_{cpu}$ need only the local $N_{cpu}$ elements of $m$ for the SPMV. When SPMV kernel acts on just $nnz1_{cpu}$ elements, we call it SPMV part 1. After the copy of $N_{gpu}$ elements of $m$ is complete, we will then commence SPMV part 2 on $nnz2_{cpu}$ elements which will complete the entire SPMV. We perform the same for $nnz_{gpu}$.
So, through this further local decomposition, we are able to achieve better overlap of computations with communication. In effect, we have achieved the 2-D decomposition of the matrix $A$.

\begin{figure}[htbp]
\includegraphics[width=\columnwidth]{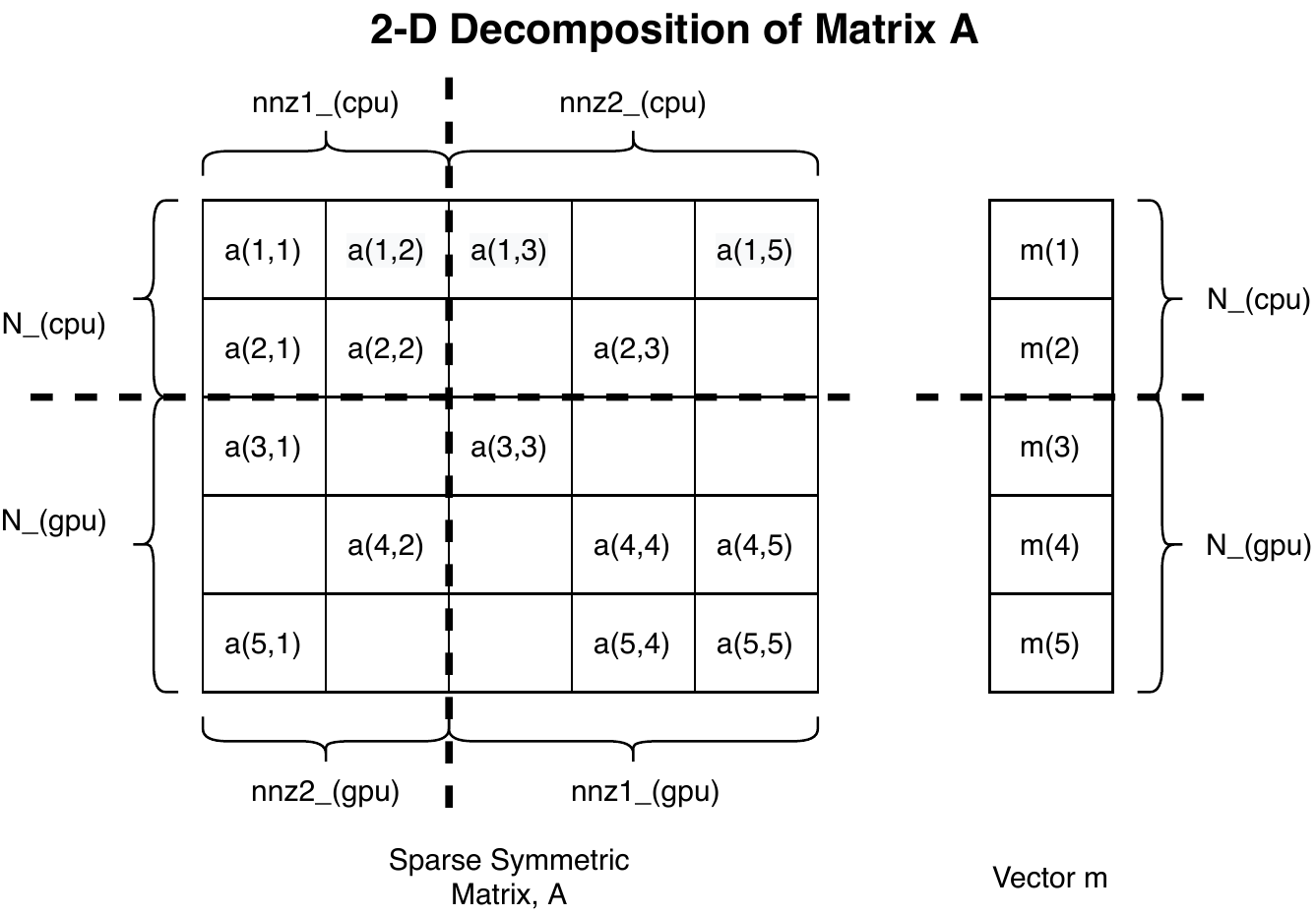}
\caption{2-D decomposition of Matrix A}
\label{matrixfig}
\end{figure}

In Figure \ref{matrixfig}, we have a matrix $A$ with $N=5$ and $nnz=15$. We see that the matrix $A$ and vector $m$ have been divided according to $N_{cpu}=2$ and $N_{gpu}=3$. We can see that $a(1,1), a(1,2), a(2,1), a(2,2)$ need only $m(1)$ and $m(2)$ for multiplication. So we assign $a(1,1), a(1,2), a(2,1), a(2,2)$ to $nnz1_{cpu}$. The rest of $nnz_{cpu}$ which need rest of the $m$ vector elements are assigned to $nnz2_{cpu}$.
We do the same for GPU and achieve the decomposition illustrated in Figure \ref{matrixfig}.

\subsubsection{Execution Flow of Hybrid-PIPECG-3 method}
\label{hybrid3flow}
\begin{figure}[htbp]
\includegraphics[width=\columnwidth]{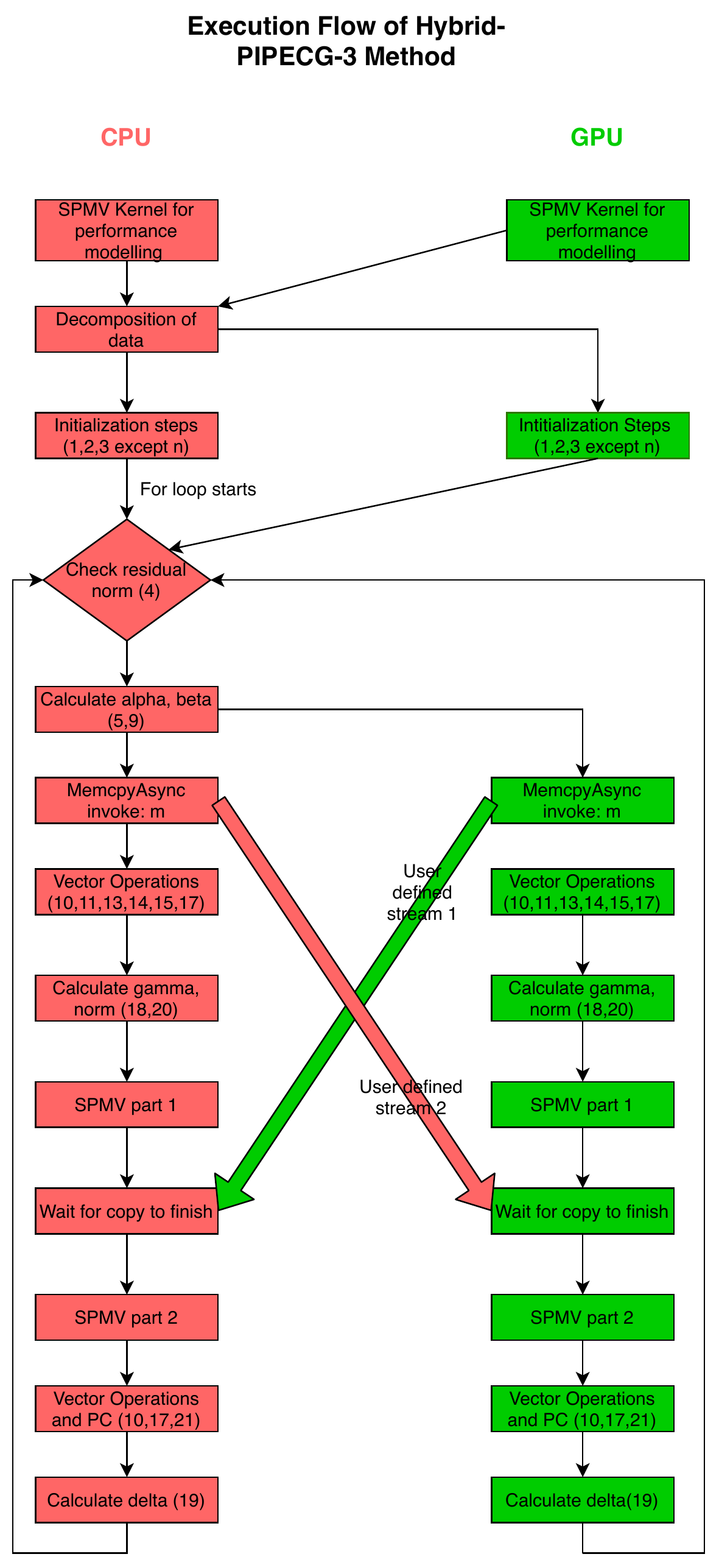}
\caption{Execution flow of Hybrid-PIPECG-3 method}
\label{Hybrid3fig}
\end{figure}

Figure \ref{Hybrid3fig} shows the execution flow of the Hybrid-PIPECG-3 method. 

For Performance Modelling, we execute the SPMV kernel on CPU and GPU simultaneously. After we get $N_{cpu}$ and $N_{gpu}$, we perform 2-D decomposition of the matrix $A$ and also decompose the vectors.

After the decomposition step, the PIPECG method starts. Both CPU and GPU perform the initialization steps on their data except the computation of $n$ vector. Then the for loop starts. After checking the residual norm, CPU calculates $\alpha$ and $\beta$. Then asynchronous copy of $m$ vector is started from CPU to GPU as well as GPU to CPU. These two Copy's are executed simultaneously using two user defined streams, Stream 1 and Stream 2. 
Similar to Hybrid-PIPECG-2 method, while CPU and GPU wait for $m$ vector to be copied so that they can calculate vector $n$, the vectors that do not depend on $n$ can be updated. This results in vector operations for $q$, $s$, $p$, $x$ and $r$. After these vector updates, $\gamma$ and  $norm$ can be computed. 
To further use the waiting time, CPU and GPU can compute SPMV part 1 as described in section \ref{data_decomposition}. 
Both CPU and GPU then wait for the Copy's to finish. With proper data decomposition, this wait is negligible as the data movement time is completely overlapped with useful computations. 
Then, CPU and GPU execute SPMV part 2 and obtain the vector $n$. They update the vectors that depend on $n$ and apply PC. Finally, they compute $\delta$ and follow the same steps iteratively.

Thus, with Hybrid-PIPECG-3, we achieve data parallelism by decomposing data between CPU and GPU. Proper overlap of the simultaneous operations executing on CPU and GPU is achieved due to decomposition of the data using the relative speeds of CPU and GPU.
Within an iteration of Hybrid-PIPECG-3, the data movement is executed through user defined streams while both CPU and GPU are busy with their useful computations that do not depend on the data that is being copied. These independent computations are provided in-part by the PIPECG method (the vector updates that do not depend on $n$) and in-part by our 2-D decomposition.

\section{Implementation and Optimizations}
\label{optimization}

\subsection{Implementation Details}
There are many preconditioners available for the CG method \cite{10.1007/978-3-319-17248-4_4} \cite{Naumov11incomplete-luand} \cite{7013063}.
Considering the time required for setting up a preconditioner and applying it in every iteration of the CG method, it is observed that iterative solvers using simple diagonal preconditioners like Jacobi preconditioner satisfactorily lower the condition number of the system and introduce less overhead \cite{8638150}. Hence, we use Jacobi preconditioner for all the methods.
Many sparse matrix formats have been proposed for getting optimum SPMV performance \cite{article}\cite{5481803} on the GPU. However, the time for converting from Compressed Sparse Row (CSR) format to these formats is significant and hence, we use CSR format.

For SPMV on the GPU, we use cusparse library's implementation as it is the most optimized for Nvidia GPU. Likewise, for dot product computations on the GPU, we use cublas library's implementation.
For PC and VMA operations on the GPU, we implement fused kernels as discussed in the next section. 

In order to efficiently use all the CPU cores present within the GPU accelerated architecture, we use OpenMP threads. We implement all the kernels of PIPECG on CPU using OpenMP constructs.

\subsection{Optimizations}
In this section, we give details of the various optimizations we have implemented for GPU and CPU.

\subsubsection{GPU related optimization}
In the PIPECG method shown in Algorithm \ref{pipelined_pcg_algo}, the lines 10-17 contain Vector-Multiply-Add (VMA) operations. In the unoptimized implementation of PIPECG, we call individual \textit{scale} and \textit{cublas daxpy} kernels to implement these operations. 
So, one operation is applied to a vector that is loaded from the global memory on the GPU to the cached local memory of the GPU. When we launch multiple kernels for multiple operations, it results in multiple reads and writes to the global memory which is time-consuming.

\begin{figure}[htbp]
\includegraphics[width=\columnwidth]{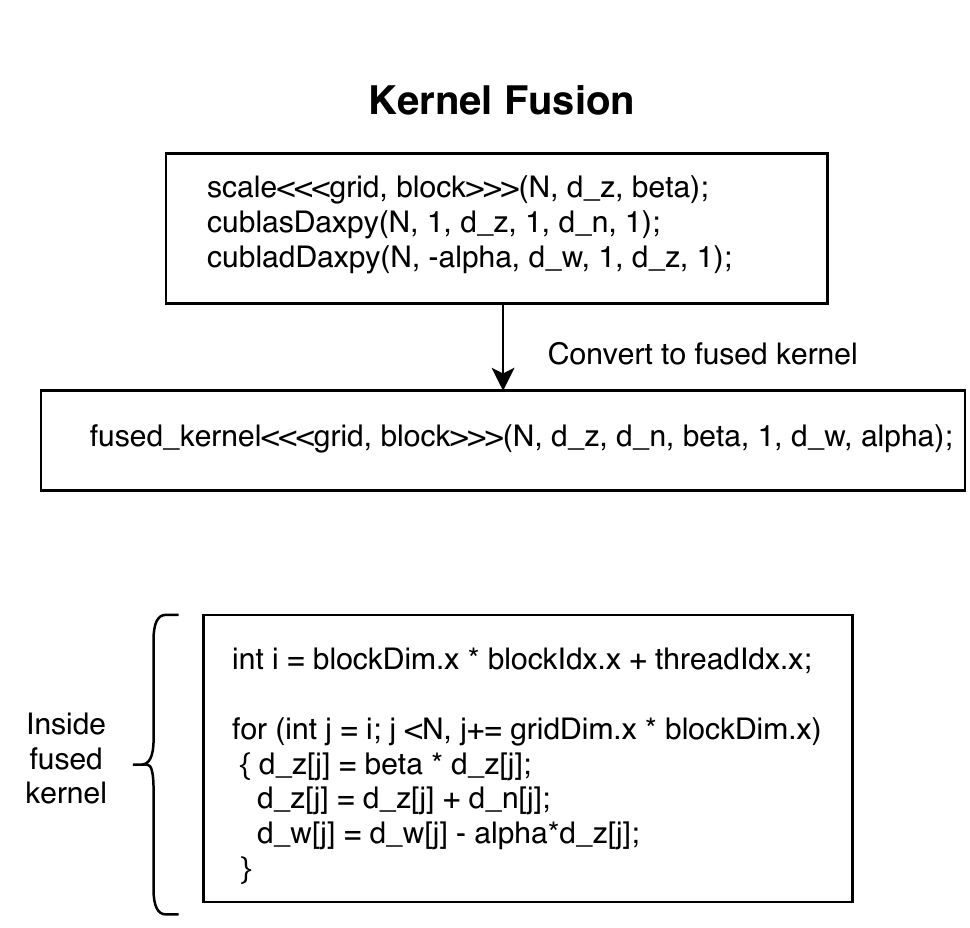}
\caption{Kernel Fusion of VMAs}
\label{kernelfusion}
\end{figure}

However, if these individual kernels are fused together into one kernel, then multiple operations can be applied to a vector loaded from the global memory to the cached local memory. This is illustrated in Figure \ref{kernelfusion}. We can see that different kernels are invoked which use $d\_z$. This results in multiple reads and writes to the global memory for $d\_z$. However, with fused kernel, $d\_z$ is only loaded once to the cached local memory and multiple operations can be performed on it.

We also fuse the Jacobi preconditioner kernel with the VMAs' fused kernels as the PC kernel uses the same vector ($d\_w$) updated by the VMAs' fused kernel.

\subsubsection{CPU related optimization}
Much like the GPU, on CPU also, the $pragma for$ loops used for implementing the VMAs can be merged so that the vectors are loaded in the cache once from the main memory and then used multiple times. This is especially beneficial for PIPECG method, as this optimization reduces the overhead introduced by the extra VMA operations in this method.

\section{Experiments and Results}
\label{experiment}
\textbf{Experimental Setup}- We run our tests on a Tesla K20m GPU with 13 Streaming Multiprocessors, 5GB memory and 16 core Intel CPU. We use 16 OpenMP threads. We run experiments on matrices from the SuiteSparse Matrix Collection\cite{Suite} as well our own generated Poisson matrices. We solve a linear system of equations $Ax=b$ with the exact solution ${x_0}=1/\sqrt{N}$, where N is the number of rows of A and $b=Ax_0$. We set the absolute tolerance to $10^{-5}$ and maximum number of iterations to 10000. We run all tests to convergence and compare the total execution times.

\subsection{Matrices from SuiteSparse Matrix Collection}
The matrices that we use from the SuiteSparse Matrix Collection are shown in Table \ref{tab1}. The first column specifies the name of the matrix, followed by the number of rows in the matrix (N) and then the number of non-zeroes in the matrix (nnz). The last column shows the ratio nnz/N. The matrices are displayed in increasing order of N.

\begin{table}[htbp]
\begin{center}
\caption{Matrices from the SuiteSparse Matrix Collection}
\label{tab1}
\begin{tabular}{|p{0.65in}|p{0.4in}|p{0.7in}|p{0.4in}|}
\hline
Matrix & N & nnz & nnz/N\\
\hline
bcsstk15 & 3948	& 117816 & 29.84\\
gyro & 17361 & 1021159	& 58.81 \\
boneS01	& 127224 & 6715152 & 52.78\\
hood & 220542 & 10768436 & 48.82\\
offshore & 259789 & 4242673	& 16.33 \\
Serena & 1391349 & 64531701	& 46.38 \\
Queen\_4147 & 4147110 & 329499284 & 79.45\\
\hline
%\multicolumn{4}{l}{$^{\mathrm{a}}$Sample of a Table footnote.}
\end{tabular}
\end{center}
\end{table}

 We compare the performance of our Hybrid-PIPECG-1, Hybrid-PIPECG-2 and Hybrid-PIPECG-3 methods with the PCG CPU and GPU implementations in the widely used Paralution\cite{paralution} and PETSc\cite{petsc-efficient} libraries. We also compare our methods with the CPU and GPU implementations of PIPECG method. We set the preconditioner to Jacobi preconditioner for all the methods. Here, we note that the total execution time for the Hybrid-PIPECG-3 method always includes the time consumed for performance modelling and 2-D data decomposition.

\begin{figure}[htbp]
\includegraphics[width=\columnwidth]{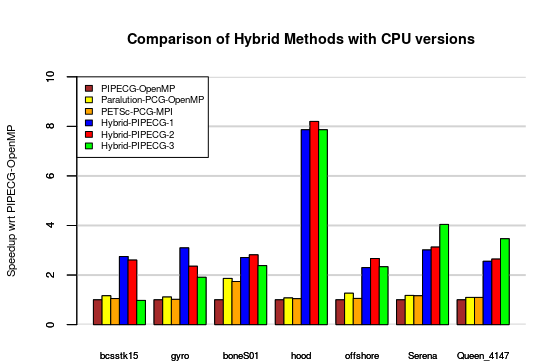}
\caption{Comparison of Hybrid methods with various CPU versions. Speedup presented wrt PIPECG-OpenMP.}
\label{cpu_comparison}
\end{figure}

\textbf{Figure \ref{cpu_comparison}} compares the performance of our hybrid methods with CPU implementations of PCG in Paralution and PETSc, and with our CPU implementation of PIPECG method.  Paralution's implementation of PCG and our implementation of PIPECG on CPU use OpenMP threads, while PETSc's implementation of PCG on CPU uses MPI processes. Our implementation of PIPECG on CPU uses OpenMP.
We present the speedups obtained by each method wrt to our PIPECG-OpenMP implementation. We observe that PIPECG-OpenMP performs the worst for every matrix. This is because the PIPECG method introduces extra VMAs to remove the dependencies. This VMA overhead is less pronounced for distributed memory systems but more pronounced for multi-core CPU in a single node. We see that PETSc-PCG-MPI always performs worse than Paralution-PCG-OpenMP. Finally, we observe that our hybrid methods perform better than all the CPU versions for all matrices because we use GPU cores as well. 

For bcsstk15 and gyro, among the hybrid methods, Hybrid-PIPECG-1 performs the best. The same behaviour is observed for matrices with N from 100 to 36000. The other hybrid methods don't perform well for these matrices with small N as Hybrid-PIPECG-2 has redundant computations on the CPU cores and Hybrid-PIPECG-3 has extra overhead of performance modelling and data decomposition.  

For boneS01, hood and offshore, Hybrid-PIPECG-2 performs the best. The same behaviour is observed for matrices with N form 36000 to 260,000. 
Hybrid-PIPECG-1 doesn't perform well for larger matrices because copying 3N elements becomes costly for large N. On the other hand, Hybrid-PIPECG-2 copies only N elements. Hybrid-PIPECG-3 performs worse than Hybrid-PIPECG-2 because in Hybrid-PIPECG-2, the vector copy is overlapped by the full SPMV kernel, whereas in Hybrid-PIPECG-3 method, it is overlapped by only SPMV part 1 kernel.

For Serena and Queen\_4147, Hybrid-PIPECG-3 performs the best.
Similar behavior is observed for matrices with N from 260,000 to 4M. Hybrid-PIPECG-1 copies 3N elements in every iteration and hence performs poorly for matrices with very large N. Hybrid-PIPECG-2 copies N elements but performs redundant computations on CPU and GPU which provide great overhead for very large N.
So, for very large N (and consequently large nnz), Hybrid-PIPECG-3 is the best suited because it provides almost perfect overlap of operations on the CPU and GPU due to our 2D data decomposition using performance modelling and asynchronous data movements.
Thus, we find that our different hybrid CPU-GPU methods give the best performance for different matrix size ranges.
\begin{figure}[htbp]
\includegraphics[width=\columnwidth]{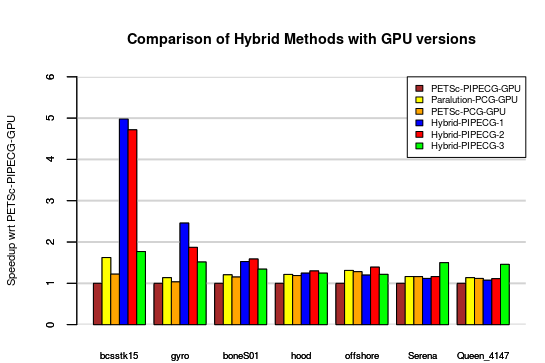}
\caption{Comparison of Hybrid methods with various CPU versions. Speedup presented wrt PETSc-PIPECG-GPU.}
\label{gpu_comparison}
\end{figure}

\textbf{Figure \ref{gpu_comparison}} compares the performance of our hybrid methods with GPU implementations of PCG in Paralution, PCG in PETSc and PIPECG method in PETSc. We present the speedups obtained by each method wrt to PETSc-PIPECG-GPU implementation. We observe that PETSc-PIPECG-GPU performs the worst for every matrix. This is because the PIPECG method introduces extra VMAs and is also not efficiently implemented for GPU on PETSc. We see that PETSc-PCG-GPU always performs worse than Paralution-PCG-GPU. Finally, we observe that our hybrid methods perform better than the the GPU versions for most of the matrices because we use multi-core CPU as well.
For offshore, Serena and Queen\_4147 matrices, we see that Paralution-PCG-GPU and PETSc-PCG-GPU perform better than Hybrid-1-PIPECG and Hybrid-2-PIPECG. This happens because as N increases, the cost of moving 3N or N elements from the GPU to the CPU in each iteration also increases. Hence, decomposing the data between CPU and GPU helps here and Hybrid-PIPECG-3 performs the best.

\subsection{125-pt Poisson Matrices}

In the previous section, we have seen the performance of our methods and Paralution and PETSc libraries matrices from the smallest N to the largest N and nnz that can be fit in our GPU's memory. Queen\_4147 (4M rows and 360M nnz's) is the largest matrix size that we are able to run on a single GPU. 
In this section, we analyse matrices that cannot be fit in the GPU memory. 
The GPU only PCG algorithms obviously can't be used for these cases. Hence, Paralution and PETSc cannot be used. Hybrid-PIPECG-1 and Hybrid-PIPECG-2 methods launch the SPMV kernel on only the GPU and thus require the full matrix to be on the GPU. Hence, these methods also cannot be used for these cases.   
In the Hybrid-PIPECG-3 method, we decompose data between CPU and GPU and achieve good performance by efficient data movement hiding and complete overlap of kernels on CPU and GPU. 

We can apply the Hybrid-PIPECG-3 method to matrices that cannot be fit into the GPU as we already have a data decomposition step in the method. Still, these matrices pose a challenge for the performance modelling step of the method. In performance modelling step, we calculate the relative speed of CPU and GPU by executing the SPMV operation on the CPU and GPU. This requires complete matrix to be on the GPU which is not possible. 

For such matrices, we run the SPMV operation on a subset $N_{pf}$ of the total number of rows in the matrix such that the nnz's contained in these $N_{pf}$ rows can be accommodated in the GPU and such that those $N_{pf}$ rows have nnz's distributed in such a way that represents the total nnz distribution of the matrix $A$. The SPMV operation is executed for the same $N_{pf}$ rows on the CPU. This would give us the required inputs for the performance modelling step. 
Ideally, we apply a heuristic to figure out such $N_{pf}$ rows. But for preliminary testing of the performance of Hybrid-PIPECG-2 method on very large matrices, we take the first N rows which contain the largest nnz that the GPU can contain.  

As the largest sparse symmetric positive definite matrix in the SuiteSparse Matrix collection is the Queen\_4147 matrix, we generate our own large matrices. We generate Poisson matrices of various sizes using a 125 point stencil as shown in Table \ref{tab2}.

\begin{table}[htbp]
\begin{center}
\caption{125 point Poisson matrices}
\label{tab2}
\begin{tabular}{|p{0.65in}|p{0.4in}|p{0.7in}|p{0.4in}|}
\hline
Matrix & N & nnz & nnz/N\\
\hline
4.5M Poisson & 4492125	& 549353259 & 122.29\\
5M Poisson & 4913000 & 601211584 & 122.37\\
6M Poisson & 5929741 & 726572699 & 120.55  \\
6.3M Poisson & 6331625 & 776151559 & 122.58 \\
\hline
%\multicolumn{4}{l}{$^{\mathrm{a}}$Sample of a Table footnote.}
\end{tabular}
\end{center}
\end{table}

\begin{figure}[htbp]
\includegraphics[width=\columnwidth]{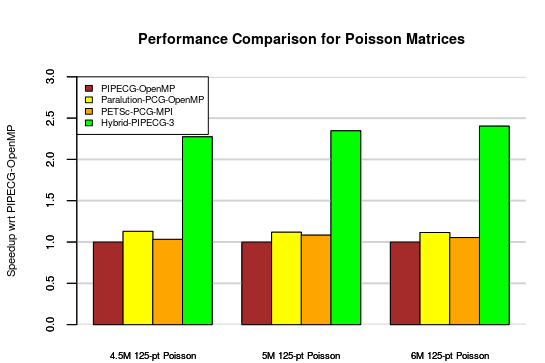}
\caption{Comparison of Hybrid-PIPECG-3 method with CPU versions for various Poisson problems. Speedup presented wrt PIPECG-OpenMP.}
\label{poissonfigure}
\end{figure}

We compare Hybrid-PIPECG-3 with CPU-only implementations of our PIPECG, PETSC PCG and Paralution PCG methods. Figure \ref{poissonfigure} shows the performance of the Hybrid-PCG and Hybrid-PIPECG-3 methods on the matrices shown in Table \ref{tab2}. 

We see that Hybrid-PIPECG-3 method performs better than PCG and PIPECG methods from Paralutiona and PETSc in all cases.
It gives 2.25 speedup for 4.5M Poisson Problem, 2.45x speedup for 5M Poisson Problem, 2.5x speedup for 6M Poisson Problem.
Thus, we find that our Hybrid-PIPECG-3 method gives 2-2.5 times speedup over the other methods.
We will get better performance improvements from Hybrid-PIPECG-3 method if use a heuristic to figure out $N_{pf}$ for performance modelling.

\section{Conclusion and Future Work}
\label{conclusion}
In this work, we proposed three methods for efficient execution of PIPECG method for GPU accelerated architecture. 
The independent computations provided by the PIPECG method gives us an opportunity for asynchronous executions on CPU and GPU. Due to asynchronous data movement, we get the benefit of both processing entities without any extra overhead from data movement.

The first method, Hybrid-PIPECG-1, achieves task parallelism by executing dot products on the CPU while GPU executes PC and SPMV kernels. This requires copy of 3N elements in every iteration. 
The second method, Hybrid-PIPECG-2 achieves task-parallelism in the same way as Hybrid-PIPECG-1. It requires copy of N elements every iteration. 
The third method, Hybrid-PIPECG-3 achieves data parallelism by decomposing the workload between CPU and GPU based on a performance model. 
The performance model takes into account the relative performance of each processing entity in heterogeneous architecture using some initial executions and performs 2D data decomposition. 

We also implement optimization strategies like kernel fusion for GPU and merging vector operations for CPU. Our methods give up to 8x speedup and on average 3x speedup over PCG CPU implementation of Paralution and PETSc libraries.  Our methods give up to 5x speedup and on average 1.45x speedup over PCG GPU implementation of Paralution and PETSc libraries. Hybrid-PIPECG-3 method also provides an efficient solution for solving problems that cannot be fit into the GPU memory and gives up to 2.5x speedup for such problems.

In the future, we plan to develop a heuristic for the performance modelling of problems that cannot be fit in the GPU. 
We also plan to extend this single node single GPU work to multiple nodes with multiple GPUs.

\bibliographystyle{IEEEtran}
\bibliography{reference}

\begin{thebibliography}{10}
\providecommand{\url}[1]{#1}
\csname url@samestyle\endcsname
\providecommand{\newblock}{\relax}
\providecommand{\bibinfo}[2]{#2}
\providecommand{\BIBentrySTDinterwordspacing}{\spaceskip=0pt\relax}
\providecommand{\BIBentryALTinterwordstretchfactor}{4}
\providecommand{\BIBentryALTinterwordspacing}{\spaceskip=\fontdimen2\font plus
\BIBentryALTinterwordstretchfactor\fontdimen3\font minus
  \fontdimen4\font\relax}
\providecommand{\BIBforeignlanguage}[2]{{%
\expandafter\ifx\csname l@#1\endcsname\relax
\typeout{** WARNING: IEEEtran.bst: No hyphenation pattern has been}%
\typeout{** loaded for the language `#1'. Using the pattern for}%
\typeout{** the default language instead.}%
\else
\language=\csname l@#1\endcsname
\fi
#2}}
\providecommand{\BIBdecl}{\relax}
\BIBdecl

\bibitem{Hestenes&Stiefel:1952}
M.~R. Hestenes and E.~Stiefel, ``Methods of conjugate gradients for solving
  linear systems,'' \emph{Journal of research of the National Bureau of
  Standards}, vol.~49, pp. 409--436, 1952.

\bibitem{10.5555/829576}
Y.~Saad, \emph{Iterative Methods for Sparse Linear Systems}, 2nd~ed.\hskip 1em
  plus 0.5em minus 0.4em\relax USA: Society for Industrial and Applied
  Mathematics, 2003.

\bibitem{10.1007/978-3-319-17248-4_4}
E.~Phillips and M.~Fatica, ``A cuda implementation of the high performance
  conjugate gradient benchmark,'' in \emph{High Performance Computing Systems.
  Performance Modeling, Benchmarking, and Simulation}, S.~A. Jarvis, S.~A.
  Wright, and S.~D. Hammond, Eds.\hskip 1em plus 0.5em minus 0.4em\relax Cham:
  Springer International Publishing, 2015, pp. 68--84.

\bibitem{Naumov11incomplete-luand}
M.~Naumov, ``Incomplete-lu and cholesky preconditioned iterative methods using
  cusparse and cublas,'' 2011.

\bibitem{7013063}
I.~{Yamazaki}, S.~{Rajamanickam}, E.~G. {Boman}, M.~{Hoemmen}, M.~A. {Heroux},
  and S.~{Tomov}, ``Domain decomposition preconditioners for
  communication-avoiding krylov methods on a hybrid cpu/gpu cluster,'' in
  \emph{SC '14: Proceedings of the International Conference for High
  Performance Computing, Networking, Storage and Analysis}, 2014, pp. 933--944.

\bibitem{article}
N.~Bell and M.~Garland, ``Efficient sparse matrix-vector multiplication on
  cuda,'' 01 2009.

\bibitem{5481803}
M.~{Mehri Dehnavi}, D.~{Fernández}, and D.~{Giannacopoulos}, ``Enhancing the
  performance of conjugate gradient solvers on graphic processing units,'' in
  \emph{Digests of the 2010 14th Biennial IEEE Conference on Electromagnetic
  Field Computation}, May 2010, pp. 1--1.

\bibitem{10.5555/898717}
E.~D’’Azevedo, V.~Eijkhout, and C.~Romine, ``Lapack working note 56:
  Reducing communication costs in the conjugate gradient algorithm on
  distributed memory multiprocessors,'' USA, Tech. Rep., 1993.

\bibitem{10.1016/0377-0427(89)90045-9}
\BIBentryALTinterwordspacing
A.~T. Chronopoulos and C.~W. Gear, ``S-step iterative methods for symmetric
  linear systems,'' \emph{J. Comput. Appl. Math.}, vol.~25, no.~2, p.
  153–168, Feb. 1989. [Online]. Available:
  \url{https://doi.org/10.1016/0377-0427(89)90045-9}
\BIBentrySTDinterwordspacing

\bibitem{mpich}
\BIBentryALTinterwordspacing
``Mpich 3.3.3,'' 2019. [Online]. Available: \url{https://www.mpich.org/}
\BIBentrySTDinterwordspacing

\bibitem{10.1016/j.parco.2013.06.001}
\BIBentryALTinterwordspacing
P.~Ghysels and W.~Vanroose, ``Hiding global synchronization latency in the
  preconditioned conjugate gradient algorithm,'' \emph{Parallel Comput.},
  vol.~40, no.~7, p. 224–238, Jul. 2014. [Online]. Available:
  \url{https://doi.org/10.1016/j.parco.2013.06.001}
\BIBentrySTDinterwordspacing

\bibitem{doi:10.1137/110838844}
N.~Bell, S.~Dalton, and L.~N. Olson, ``Exposing fine-grained parallelism in
  algebraic multigrid methods,'' \emph{SIAM Journal on Scientific Computing},
  vol.~34, no.~4.

\bibitem{10.1007/s11227-012-0825-3}
R.~Li and Y.~Saad, ``Gpu-accelerated preconditioned iterative linear solvers,''
  \emph{J. Supercomput.}, vol.~63, no.~2, 2013.

\bibitem{10.1145/1654059.1654078}
N.~Bell and M.~Garland, ``Implementing sparse matrix-vector multiplication on
  throughput-oriented processors,'' in \emph{SC 2009}, 2009.

\bibitem{10.1145/2907944}
\BIBentryALTinterwordspacing
K.~Rupp, J.~Weinbub, A.~J\"{u}ngel, and T.~Grasser, ``Pipelined iterative
  solvers with kernel fusion for graphics processing units,'' \emph{ACM Trans.
  Math. Softw.}, vol.~43, no.~2, Aug. 2016. [Online]. Available:
  \url{https://doi.org/10.1145/2907944}
\BIBentrySTDinterwordspacing

\bibitem{https://doi.org/10.1002/fld.2462}
S.~Georgescu and H.~Okuda, ``Conjugate gradients on multiple gpus,''
  \emph{International Journal for Numerical Methods in Fluids}, vol.~64, 2010.

\bibitem{8638150}
M.~{Götz} and H.~{Anzt}, ``Machine learning-aided numerical linear algebra:
  Convolutional neural networks for the efficient preconditioner generation,''
  in \emph{2018 IEEE/ACM 9th Workshop on Latest Advances in Scalable Algorithms
  for Large-Scale Systems (scalA)}, 2018, pp. 49--56.

\bibitem{Suite}
\BIBentryALTinterwordspacing
``Suitesparse matrix collection,'' 2020. [Online]. Available:
  \url{https://sparse.tamu.edu/}
\BIBentrySTDinterwordspacing

\bibitem{paralution}
\BIBentryALTinterwordspacing
P.~Labs, ``Paralution v1.1.0,'' 2020. [Online]. Available:
  \url{http://www.paralution.com/}
\BIBentrySTDinterwordspacing

\bibitem{petsc-efficient}
S.~Balay, W.~D. Gropp, L.~C. McInnes, and B.~F. Smith, ``Efficient management
  of parallelism in object oriented numerical software libraries,'' in
  \emph{Modern Software Tools in Scientific Computing}, E.~Arge, A.~M. Bruaset,
  and H.~P. Langtangen, Eds.\hskip 1em plus 0.5em minus 0.4em\relax
  Birkh{\"{a}}user Press, 1997, pp. 163--202.

\end{thebibliography}


\begin{thebibliography}{1}

\bibitem{IEEEhowto:kopka}
H.~Kopka and P.~W. Daly, \emph{A Guide to \LaTeX}, 3rd~ed.\hskip 1em plus
  0.5em minus 0.4em\relax Harlow, England: Addison-Wesley, 1999.

\end{thebibliography}

\end{document}